# Experimental Study of Magnetic Near-Field Microstrip Electronic Probe for PCB EMC Emission Measurement

Hongchuan Jia
*NUIST*
Nanjing, China
email: 1104429556@qq.com

Fayu Wan
*NUIST*
Nanjing, China
email: fayu.wan@nuist.edu.cn

Vladimir Mordachev
*R&D Laboratory of EMC*
*BSUIR*
Minsk, Belarus
email: mordachev@bsuir.by

Jerome Rossignol
*Université de Bourgogne*
Dijon, France
email: jerome.rossignol@u-bourgogne.fr

Glauco Fontagalland
*Brenton School of Engineering*
*University of Mount Union*
Alliance, OH 44601 USA
email: fontgalland@gmail.com

Nour Mohammad Murad
*PIMENT, IUT*
*University of La Réunion*
Saint Pierre, La Reunion
nour.murad@univ-reunion.fr

Blaise Ravelo
*NUIST*
Nanjing, China
ORCID: 0000-0001-7334-5016

*Abstract*—An experimental study on magnetic near-field (NF) scanning of printed circuit board (PCB) emission radiation is developed in this paper. The design and installation of the electromagnetic (EM) NF scanner is introduced. The test bed of magnetic NF emission in the microwave frequency range is described. The methodology of the microstrip magnetic NF probe is discussed. The probe calibration process was performed following the IEC 61967-1 NF scanning standard. The NF scanner functioning is tested with passive microstrip circuit square loop probe and device under test (DUT) PCB radiation in the test plan positioned at 1-mm above the ground plane. Based on the standard test with I-shape 50-Ω transmission line (TL), the calibration process of radiated magnetic field was validated by comparison between HFSS® simulation and experimentation in very wideband frequency from 0.1-GHz to 3-GHz. Then, a non-standard TL based DUT was experimented. Accordingly, the cartographies of scanned magnetic NF at two different test frequencies, 2 GHz and 3 GHz, are discussed. The NF scanner is under development for targeting the EMC radiated emission of PCB dedicated to operate in 6G wireless communication.

I. INTRODUCTION

Nowadays, the near-future of radio frequency (RF) and microwave wireless communication engineering is expecting to be developed toward 6G technology [1]. In spite of the spectacular progress of technology design, the wireless and mobile communication system operates challenging electromagnetic background [2-3]. The design and implementation of 6G RF system necessitate some research challenges to overcome notable technological problems [4].

Therefore, one wonders nowadays "How will 6G affect electromagnetic compatibility (EMC)?" [5]. Today, such an open question becomes a challenging and hot point for both conduced and radiated EMC characterization with different hardware system level in the research communities [6-7]. So far, the radiated EMC tests constitute one of the challenging tasks for any electronic devices and systems [8]. Some EMC compliance test standards for residential electronic device as EN 61000-4-3 remain familiar and practical but notably time cost ones [3]. But for the case of planar printed circuit board (PCB) as device under test (DUT), the electromagnetic (EM) near-field (NF) scanning [9] following international standard as IEC 61967 [10] seems to be the most efficient technique.

Different research works on NF scanning with high precision measurement system [11], and for shielding effectiveness [12] were developed. The NF scanning can also be used for developing and PCB radiated emission modelling [13][14]. The performance of PCB EM NF scanning depends systematically on the used probes [15][16].

The present research work introduces an experimental study of EMC radiated emission. The feasibility study of magnetic NF scanning is based on both standard and non-standard cases of PCB as DUT.

## II. DESCRIPTION OF TEST BED DEDICATED TO MAGNETIC NF EMISSION OF PCB

This section describes the methodology of test bed installation for the EM NF emission scanning of planar PCB under investigation.

### A. Radiating PCB and Measurement Equipment Test Setup

The synoptic diagram of NF scanner testbed recently designed in the EMC lab of NUIST including the PCB DUT, probe and two-port vector network analyzer (VNA) is illustrated by Fig. 1(a). This EM NF scanning is generally used to study the radiation of electronic components and devices [10][11][12][13][16][17][18]. In this case, we consider the standard case of PCB constituted by TL with characteristic impedance equal to resistive load $Z_0$=50 Ω. For the case of magnetic NF scanning, the face view of test is depicted by Fig. 1(b).

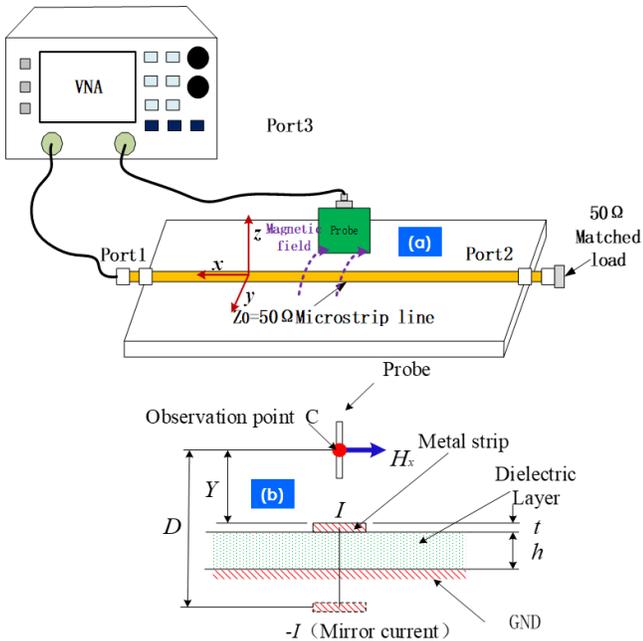

Fig. 1. (a) Illustrative diagram of EM NF scanner and (b) face view of planar PCB as DUT and its geometrical parameters.

In this test diagram, the probe is aimed to measure the $H_x(f)$ component versus frequency $f$ in the scanning plan place at the height $y$ above the DUT PCB metallization and dielectric substrate with thicknesses $t$ and $h$, respectively [10]. During the test, the PCB constituting conductor is flowed by current with frequency dependent intensity $I(f)$. The distance between the test point and the symmetrical conductor line with reference to the DUT ground plane is equal to $d$. The formula of the measured magnetic field extraction from the touchstone data of the transmission coefficient $S_{21}(f)$ is discussed in the following subsection.

### B. Magnetic NF Calibration Equation Following IEC 61967-1 Standard

It is worth to recall the analytical expression for the probe calibration before the test and result analysis. Based on the

The present research work is supported by National Natural Science Foundation of China (NSFC) Grant No. 62350610268.

Faraday's law, the frequency dependent magnetic field $H(f)$ induces voltage $V(f)$ across the SMA access port with antenna factor defined by:

$$CF(f) = \frac{H(f)}{V(f)}. \quad (1)$$

According to the standard [10], this probe antenna factor can be expressed in function of transmission coefficient $S_{31}(f)$ by:

$$CF_{dB}(f) = 20\log_{10}\left[\frac{d}{\pi h(h+2d)}\right] - S_{31dB}(f) - 34. \quad (2)$$

By using the calibration formula, the measured magnetic field components from the voltage $V_{mes}$:

$$H_{mes,dB}(f) = \begin{cases} 20\log_{10}\left[\dfrac{d}{\pi h(h+2d)}\right] - S_{31dB}(f) \\ -34 - V_{mes,dB}(f) \end{cases} \quad (3)$$

one gets the experimental results discussed in the following section.

## III. CALIBRATION METHODOLOGY OF MAGNETIC NF PROBE

The present section examines the calibrated results of magnetic NF test results following the IEC 61967 standard [10].

### A. Magnetic NF Design in Microstrip Technology

The photo of the magnetic NF microstrip probe explored in the present study is shown in Fig. 2. This magnetic probe was designed as a square loop geometrical shape constituted by TL with width $w_p$=0.5 mm and side $s$=4 mm and ended by SMA connector.

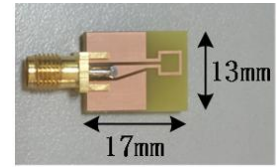

Fig. 2. Photo of magnetic NF microstrip probe.

The fabricated probe was implemented in microstrip technology. It acts as a square metallization loop passive circuit printed on FR4 substrate with relative permittivity $\varepsilon_r$, loss tangent $\tan(\delta)$, dielectric substrate thickness $h$ and Copper metallization with thickness $t$ as indicated in Table I.

TABLE I. SPECIFICATIONS OF CONSIDERED SUBSTRATE CONSTITUTING THE NF PROBE

| Description | Name | Value |
|---|---|---|
| Dielectric substrate thickness | $h$ | 1.6 mm |
| Relative permittivity | $\varepsilon_r$ | 4.6 |
| Metallization thickness | $t$ | 35 µm |
| Loss tangent | $\tan(\delta)$ | 0.016 |
| Metallization conductivity | $\sigma$ | 58 MS/s |

The test setup installation using this microstrip magnetic NF probe is presented in the next subsection.

## B. Test Setup

Fig. 3 represents the test setup of the magnetic or H-probe shown in Fig. 2 and the PCB DUT which is dedicated to sensitivity analysis [10][11][12][13][16][17][18]. It acts as manual movement test of the probe to detect its experimental sensitivity. As a standard test circuit, the DUT is constituted by "I"-TL implemented in microstrip technology on the same substrate. The tested TL characteristic impedance was chosen to be $Z_0$. Otherwise, the constituting conductor width is fixed to $w_{DUT}$=3 mm.

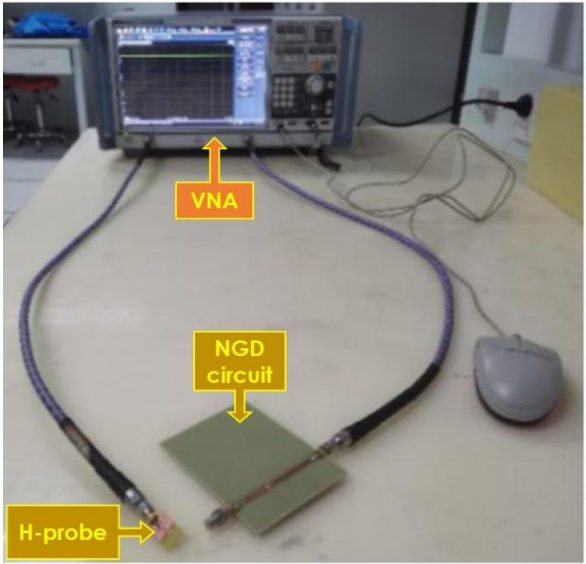

Fig. 3. Photo of experimental setup using magnetic NF microstrip probe and DUT PCB.

The proposed test is based on the S-parameter measurement as illustrated by the diagram of Fig. 1. The two-port VNA is connected to the DUT in port ❶ and to the probe in port ❷. The VNA from Rohde&Schwarz® referenced ZNB 20 and specified by frequency band 100 kHz to 20 GHz was used during the experimental test.

The obtained results of magnetic NF from the standard test are discussed in the following subsection.

## C. Results of Magnetic NF Calibration with I-Shape Microstrip Line

According to the IEC 61967 standard [10], $h$=1 mm was considered during the test.

The output power of the excitation signal delivered by VNA is -10 dBm. The results discussed in the present subsection are obtained from the test illustrated by Fig. 3 and 3-D EM simulations from HFSS® commercial tool provided by ANSYS®. The performed study was spanned in the very wide frequency range with $f_{min}$=0.1 GHz and $f_{max}$=3 GHz. The obtained results of reflection and transmission parameters are plotted by Fig. 4(a). We can find the high-pass filter behavioral response of the magnetic probe [17].

The magnetic probe antenna factor versus frequency is displayed by Fig. 4(b). We find that the simulated and measured results are in good agreement. Then, the extracted $H_y(f)$ from measurement which intensity decreases inversely to the frequency is shown by Fig. 3(c).

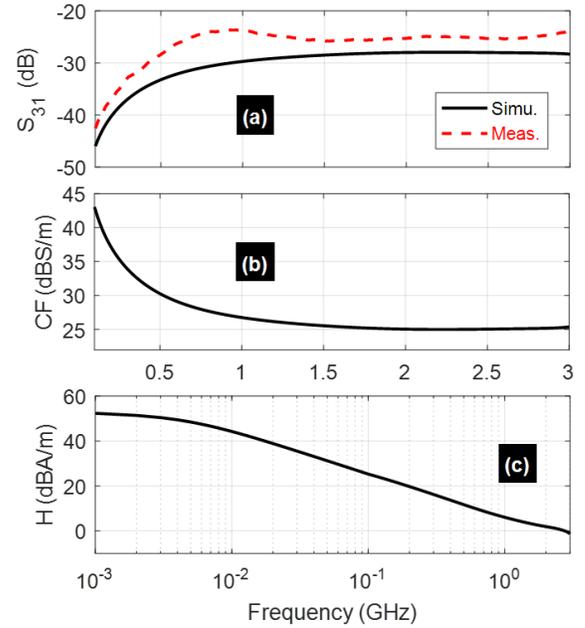

Fig. 4. Comparison of simulated and measured (a) S-parameters, (b) probe antenna factor and (c) magnetic NF versus frequency.

To confirm the performance of the designed magnetic NF probe, EMC radiations of DUTs were tested. The following section will discuss on the 1-D and 2-D plotting of tested PCB emissions.

## IV. DISCUSSION ON THE MEASURED RESULTS OF SCANNED PCB MAGNETIC NF

The present section examines the magnetic NF test results scanned from PCB radiated emission. One emphasizes that the magnetic NF was measured along a rectangular surface delimited by $x_{min}$, $x_{max}$, $y_{min}$ and $y_{max}$ with resolutions $\Delta x$ and $\Delta y$, as illustrated in Fig. 5.

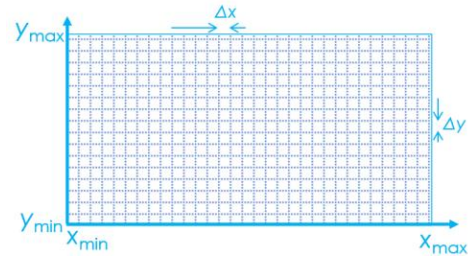

Fig. 5. Scanning surface representation.

### A. 1-D Plotting of Magnetic NF from Standard Microstrip DUT

The comparison of simulated and measured profiles of $H_y(f_1$=2 GHz) from the test diagram illustrated by Fig. 1 along y-axis with $y_{min}$ and $y_{max}$ under $\Delta y=\Delta x$ addressed in Table II is plotted in Fig. 5.

TABLE II. SCANNING SURFACE GEOMETRICAL PARAMETERS AND MIN/MAX OF SCANNED MAGNETIC NF

| Nature | Name | Value |
|---|---|---|
| Geometrical parameter | $y_{min}$ | -5 mm |
| | $y_{max}$ | 5 mm |
| | $\Delta y$ | 0.5 mm |
| Y-component of scanned magnetic NF | $H_{y,min}$ | -20 dBA/m |
| | $H_{y,max}$ | 0 dBA/m |

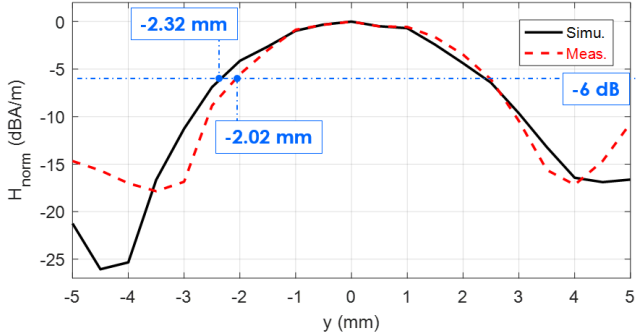

Fig. 6. Profiles comparison of magnetic NF radiated by DUT PCB along y-axis.

Once again, the simulated and experimented magnetic fields which vary from about $H_{y,min}$ and $H_{y,max}$ which are addressed in Table II are in good agreement.

The magnetic NF profile results which confirm the feasibility study of the test bench led us to the mapping plot of the next subsection.

### B. Discussion on Non-Standard Magnetic NF Measured Results

In this case of study, a non-standard DUT of microstrip circuit shown by Fig. 6 was tested. This DUT was implemented on the FR4 substrate similar to the NF probe one specified by Table I. The scanning test was carried out following the testbed described earlier in Fig. 1.

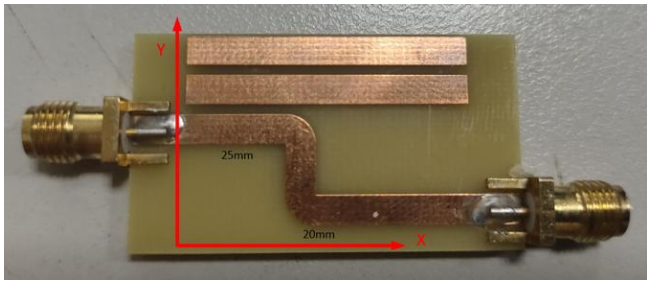

Fig. 7. Photo of tested non-standard DUT PCB.

The scanning plan was the horizontal surface with size $L_x \times L_y$ under resolution $\Delta x = \Delta y$ indicated in Table III which was placed at the height $L_z = 1$ mm above the metallic plane of the DUT. After the application of calibration equation (3), we obtain $H_y(f_1=2$ GHz) and $H_y(f_2=3$ GHz). The corresponding cartographies are mapped by Fig. 7(a) and Fig. 7(b), respectively.

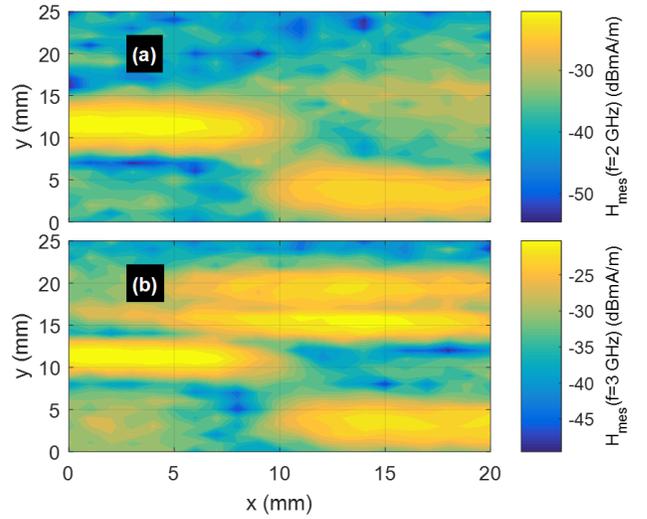

Fig. 8. Measured magnetic NF test radiated by the PCB shown in Fig. 6 at (a) $f_1=2$ GHz and (b) $f_2=3$ GHz.

TABLE III. SCANNING SURFACE GEOMETRICAL PARAMETERS AND MIN/MAX OF SCANNED MAGNETIC NF

| Nature | Name | Value |
|---|---|---|
| Geometrical parameter | $L_x$ | 20 mm |
| | $L_y$ | 25 mm |
| | $\Delta x = \Delta y$ | 0.5 mm |
| Y-component of scanned magnetic NF | $H_{y,min}(f_1)$ | -53 dBA/m |
| | $H_{y,max}(f_1)$ | -21 dBA/m |
| | $H_{y,min}(f_2)$ | -49 dBA/m |
| | $H_{y,max}(f_2)$ | -21 dBA/m |

The cartographies of measured magnetic NF highlight the occurrence of stronger EM radiation with minimum and maximum values, $H_{y,min}$ and $H_{y,max}$, respectively, shown in Table III above the metallized zone of the conductor. The EM NF signature of the DUT is completely different for the tested two frequencies. At $f=2$ GHz, the cartography does not enable to see the tested PCB interspace. However, the higher test frequency $f=3$ GHz, one observed a clear variation of magnetic NF corresponding to the interspace pattern of the DUT. In the continuation of the present study, we are planning to study the EM NF emitted by millimeter-wave electronic devices.

## V. CONCLUSION

Among the different measurement methods, the NF scanning is one of fasted radiated EMC test techniques. However, the NF scanning is still not familiar to most of electronic device designer and manufacturers. For this reason, a practical study highlighting the basic steps of PCB EMC NF scanning is still necessary.

An experimental study of microstrip probe for EM NF scanning is reported. The proposed study is dedicated to planar PCB EMC emission. The methodology of the NF scanning following the IEC 61967-1 standard calibration is described. Results of illustrative test case with microstrip magnetic NF probe and passive RF circuit as DUTs with 1-D and 2-D plotting are discussed.

In the continuation of the present work, the development of NF scanner dedicated to 6G PCBs and electric fast transient event notably in the time-domain as introduced in [17][18] is currently in progress.

ACKNOWLEDGMENT

The present research work is supported by National Natural Science Foundation of China (NSFC) Grant No. 62350610268.